\documentstyle[prb,aps]{revtex}

\def\beqn{\begin{eqnarray}}
\def\eeqn{\end{eqnarray}}
\def\beqns{\begin{eqnarray*}}
\def\eeqns{\end{eqnarray*}}
\def\beq{\begin{equation}}
\def\eeq{\end{equation}}
\def\bea{\begin{array}}
\def\ea{\end{array}}

\def\<{\langle}
\def\>{\rangle}

\title{Electronic transport properties of quasicrystals}

\author{S. Roche, G. Trambly de Laissardiere and D. Mayou}

\begin{document}

\maketitle

\begin{center}
Laboratoire d'Etudes des Proprietes Electroniques des
Solides-CNRS, Grenoble
\end{center}

\begin{abstract}
\leftskip 54.8pt
\rightskip 54.8pt
We present a review of some results concerning electronic transport
properties of quasicrystals. After a short introduction to the basic
concepts of quasiperiodicity, we consider the experimental transport
properties of electrical conductivity with particular focus on the effect
of temperature, magnetic field and defects. Then, we present some heuristic
approaches that tend to give a coherent view of different, and to some
extent complementary, transport mechanisms in quasicrystals. Numerical
results are also presented and in particular the evaluation of the linear
response {\it Kubo-Greenwood} formula of conductivity in quasiperiodic
systems in presence of disorder.
\end{abstract}

\vspace{20pt}


\newpage

\tableofcontents

\newpage

\section{INTRODUCTION}

\vspace{20pt}

\hspace{\parindent}In 1984, D. Schechtman, I. Blech, D. Gratias and J.W.
Cahn \cite{AlMn} presented a new metastable phase of an AlMn binary alloy.
The diffraction pattern was formed by intense Bragg peaks organized
according to the icosahedral symetry strictly forbidden from conventional
crystallography. The underlying order was claimed to be described by the
mathematical concept of quasiperiodicity \cite{Bohr,SteinI}.

\vspace{10pt}

\hspace{\parindent}The confirmation of a new state of matter has been an
intense subject of controversy. In particular L. Pauling proposed
alternative description of five-fold diffraction patterns based on
icosahedral glasses formed by twins \cite{Pauling}. However, the situation
changed after the discovery of stable phases (icosahedral $AlCuFe$,
$AlPdMn$, $AlCuCo$...) by Tsai et al. \cite{Tsai}, and the existence of
quasiperiodic crystals (quasicrystals) is now well accepted. Furthermore,
these materials have revealed a lot of unexpected physical properties
\cite{Avignon}.

\vspace{10pt}

\hspace{\parindent}The aim of this article is to review briefly the
experimental results concerning electronic conductivity and then to present
theoretical studies of electronic structure and electronic transport in
these systems.

\vspace{10pt}

\section{Quasiperiodic order}

\vspace{10pt}

\subsection{Construction of the Fibonacci chain}

\vspace{10pt}

\hspace{\parindent}To generate the Fibonacci chain, it is possible to use
an inflation process starting from two incommensurate segments of
respective lengthes A and B. One adopts the rule $\hbox{A}\to \hbox{AB}$,
and $\hbox{B} \to \hbox{A}$,  which leads to the consecutive sequences
{\small $S_{0}=\hbox{A}$},{\small  $S_{1}=\hbox{AB}$}, {\small
$S_{2}=\hbox{ABA}$}, {\small $S_{3}=\hbox{ABAAB}$} {\small
$S_{4}=\hbox{ABAABABA}$}, {\small $S_{5}=\hbox{ABAABABAABAAB}$} and so on.
This inflation rule is related to the sequence of Fibonacci numbers given
by

\vspace{5pt}

$$F_{n}=F_{n-1}+F_{n-2}, F_{0}=0, F_{1}=1 \ \ \hbox{with
} \ \  F_{n+1}/F_{n} \to \tau=(1+\sqrt{5})/2=1.61803398...$$

\vspace{10pt}

\noindent
In the infinite limit, the ratio of numbers of A and B of the semi-infinite
chain \\
{\small $S_{\infty}=\hbox{ABAABABAABAABABAABABA}...$} is equal to the
golden mean $\tau$.

\vspace{10pt}

\hspace{\parindent}The cut and project method \cite{DuneauI} is
 a geometrical algorithm to generate quasiperiodic structures in a D
dimensional
space starting from a periodic one in N dimensional space (with
$N>D$). It can be
 related to the fact that every quasiperiodic function can be
algebraically
 related to a periodic one in higher space. If one defines $f(x,y)$ a
$2\pi
 $-periodic function in  $x$ and $y$ directions, Fourier decomposition
of this function
 leads to $f(x,y)=\sum_{p,q}\alpha_{pq}\exp(2i\pi(px+qy))$.
The quasiperiodic function $g(x)=f(x,y=x/\tau)$ in 1D space can be
seen as a restriction
 to $y=x/\tau$ (line with irrational slope) of
a periodic function in 2D space. We illustrate this cut and project method
for the construction of the Fibonacci chain in figure figFCP where two main
steps are distinguishables.

\vspace{10pt}

\hspace{\parindent}Firstly, one defines a window of width ${\cal B}$ along
a line  ${\cal D}$ of slope p. Next, the vertices of the square lattice in
2D that belong to the window, are projected perpendicularly on ${\cal D}$.
These projected points define a sequence of two distinct lengthes A and B
such that $A/B=p$. When p is rational, the chain is periodic, whereas for
irrational slope, it becomes quasiperiodic. The Fibonacci chain correspond
to a slope $p=1/\tau$ (where $\tau$ is the golden mean). The periodic
approximant chains of Fibonacci are related to the sequence of rational
numbers $p_{n}$ converging to the slope that determines the quasicrystal.
For example, for the Fibonacci chain they are defined by
(1/1,2/1,3/2,5/3,...$p_{n}=F_{n-1}/F_{n}\to \tau$). So, they do not share
long range quasiperiodic order in a strict sense, but are close from the
quasiperiodic chain as far as local order is concerned.

\vspace{20pt}

\subsection{Quasiperiodic tilings}

\vspace{20pt}

\hspace{\parindent}The possibility of quasiperiodic tilings of an
euclidian space was known since the work of R. Penrose
\cite{Penrose}. On figure fig2  we represent a part of a Penrose
 tiling which is constructed from two rhombi of respective angles
 ({\small $\frac{2\pi}{5},\frac{3\pi}{5}$}) and ({\small
$\frac{\pi}{5},\frac{4\pi}{5}$})
which are assembled according to local matching rules \cite{Pqp}. The
Penrose tiling is constrained to a long range five-fold orientational
order, which corresponds to a diffraction pattern shown in figure
figcdiff.
An algebraic description of Penrose tiling is due to De Brujin \cite{Brujin}.

\vspace{10pt}

\hspace{\parindent}If we consider a tiling of a N-dimensional space
${\cal E}$ a quasiperiodic tiling of a D-dimensional space ${\cal
E}_{\parallel}$ is described in a general method developed by
M. Duneau and A. Katz \cite{DuneauK}. This method generalizes the
example of figure figFCP,
 where ${\cal E}_{\parallel}$ will contain the Fibonacci chain. The
complementary space is
 called ${\cal E}_{\perp}$, of $(N-D)$ dimensions, and such that
${\cal E}_{\parallel}+{\cal E}_{\perp}={\cal E}$. In real space ${\cal
E}_{\parallel}$
 the geometrical properties of quasiperiodic
tilings can be summarized by the followings points \cite{DuneauI,Katz,SteinM} :

\vspace{5pt}

\begin{itemize}
\item
{\bf Aperiodicity} : No discrete translation lets the tiling invariant.
\item
{\bf  Local Isomorphism} ({\bf Conway theorem}) : Every finite partition of
tiles of characteristic length ${\cal L}$ admits at least one identical
replica in a distance of order $2{\cal L}$. This frequent and regular
reappearances of identical local environments reveals a mesoscopic
homogeneity which gives an intermediate situation between periodic
translational order and disordered tilings.
\item
{\bf Self-similarity}: A tiling is said self-similar if it exists another
one composed by smaller tiles (with a ratio of length of $(1+\sqrt{5})/
2=\tau$ for a Penrose tiling) which preserves all the vertices of the
primary tiling. This property allows to define scale transformations from a
tiling to an inflated or deflated one.
\end{itemize}

\vspace{10pt}

\noindent
Physical quantities (like for example the charge density $\rho({\bf r})$)
admits Fourier components with vector ${\bf K}$ that can be indexed with
integers. For example, 3D-icosahedral quasicrystals requires the use of a
basis of six integers such that every ${\bf K}$ vector of the reciprocal
space is defined by

$${\bf K}=n_{1}{\bf a}^{*}_{1}+n_{2}{\bf a}^{*}_{2}+n_{3}
{\bf a}^{*}_{3}+n_{4}{\bf a}^{*}_{4}+n_{5}{\bf a}^{*}_{5}+
n_{6}{\bf a}^{*}_{6}$$

\vspace{5pt}

\noindent
where the ${\bf a}^{*}_{i}$ are chosen according to the five-fold axis of
an icosahedra, and every ${\bf K}$ are decomposed in an orthonormal basis
according to $(h+\tau h', k+\tau k',l+\tau l')$ with $h,h',k,k',l,l'$ some
integers and $\tau$ the golden mean.

\vspace{20pt}

\subsection{Atomic cluster models}

\vspace{30pt}

\hspace{\parindent}Once the quasiperiodic order was proposed as a candidate
to explain the long-range special symmetry inherent to quasicrystals, the
major work for crystallographers has been to develop structural models of
atomic distribution. For icosahedral phases, atomic models for alloys such
as $AlCuFe$ and $AlPdMn$ \cite{Janot2,Gratag,Ag1,Ag2} have been proposed,
with an estimated accuracy of about $80-90\%$ of the real structure.

\vspace{10pt}

\hspace{\parindent}We give on figures {Jan1} and {Jan2}, an example of a
structural model proposed by C. Janot \cite{Janot2} for the icosahedral
phase of $AlPdMn$. In figure {Jan1} the ``elementary brick" known as the
pseudo-Mackay icosahedron (P.M.I.) is used to construct the quasiperiodic
tiling according to inflation rules (some additional atoms have to be added
to fill entirely the quasiperiodic lattice).

\vspace{20pt}

\section{Experimental transport properties of quasicrystals}

\vspace{30pt}

\hspace{\parindent}Here, we will focus on electronic transport properties
as a function of temperature, magnetic field and defects concentration, and
present essentially dc conductivity results.

\vspace{10pt}

\subsection{Different classes of quasicrystals}

\vspace{20pt}

\hspace{\parindent}Quasicrystals have been divided into two main families
which are the icosahedral phases whose diffraction pattern have five-fold
symetry and which are quasiperiodic in the 3 directions of space, and
decagonal phases which confine quasiperiodicity in a 2D plane, leaving one
direction periodic. The second classe gives the possibility to compare
properties between quasiperiodic and periodic directions. For example, a
strong conductivity anisotropy between quasiperiodic planes and periodic
direction has been measured \cite{anisoI}. Yet in the following, we focus
essentially on icosahedral phases that have the most striking properties.

\vspace{20pt}

\subsection{Electronic conductivity of quasicrystals}

\vspace{20pt}

\hspace{\parindent}Conductivity in icosahedral phases of high structural
quality such as $AlCuFe$ or $AlPdMn$ alloys is very low, of the order of
$100-300 {(\Omega cm)}^{-1}$ at zero temperature. Moreover, we  illustrate
on figure {figicopar}, one of the most unexpected transport properties of
quasicrystals. Indeed, after annealing samples, with a consequent
improvement of structural quality, the conductivity decreases
\cite{Klein2,May1I}. This tendency is surprising since these phases are
generally composed of good metals for which an increase of conductivity is
expected with improvement of order. Consequently, the correlation between
transport properties and quasiperiodic order has been an intense subject of
discussion and controversy.

\vspace{10pt}

\hspace{\parindent}We also note that the curves $\sigma(T)$  are nearly
parallel. This point suggests to write
$\sigma(T)=\sigma_{4K}+\delta\sigma(T)$ with $\sigma_{4K}$ a measure  of
conductivity at $4$ Kelvin and $\delta\sigma(T)$ the variation as a
function of temperature which is nearly independent of the alloy (fig.
{figicopar}). This is a general behavior of all the icosahedral
quasicrystals of high structural quality and corresponds to an ``inverse
Mathiessen rule" \cite{May1I}. Indeed, the Mathiessen  rule
$\rho(T)=\rho_{0}+\delta\rho(T)$ is characteristic of metallic alloys where
$\rho_{0}$ and $\delta\rho(T)$  are respectively the resistivity due to
static defects and scattering by phonons.

\vspace{10pt}

\hspace{\parindent}Generally, the phase diagram of quasicrystals in
composition and temperature is very complex and the existence zone of
quasiperiodic order is reduced. Around these particular zones of the phase
diagram, some other periodic phases can be grown with a similar local
order. These so-called approximants phases possess a common local order
with quasicrystals and seem to have the same kind of physical properties.
This is the case for alloys such as ({\small $AlCuFe, AlMnSi, AlPdMn$},...)
which are associated with families of crystalline approximants phases like
{\small $\alpha-AlMnSi, \alpha-AlCuFe, R-AlCuFe...$} whose smallest unit
cell can have only $\sim 130$ atoms. However the alloy {\small $AlPdRe$ }
is an exception \cite{AvignonP}, since no approximant phases are known.

\vspace{10pt}

\hspace{\parindent}We show, in figure {fig2}, recent results where the
conductivity of a small cubic approximant {\small $\alpha-AlSiCuFe$} (with
a unit cell parameter of $a=12,33\AA$) is compared to that of an
icosahedral phase of $AlCuFe$ \cite{CB_RQ9}. The difference of the absolute
conductivity is very small and in addition, the behavior of
{\small$\delta\sigma(T)>0$} is the same. In contrast, a crystalline
``non-approximant" phase of similar stoichiometry (tetragonal {\small
$\omega-Al_{7}Cu_{2}Fe$} phase) with cell parameters $a=6,34\AA$ and
$c=14,87\AA$, has a conductivity of the order of  {\small $10^{4} (\Omega
cm)^{-1}$} between 0 and 300K (with a metallic temperature dependence
{\small$\delta\sigma(T)<0$}).

\vspace{10pt}

\hspace{\parindent}Another interesting experimental result concerning
approximants has been revealed by two successive approximants of the cubic
phases of {\small $AlGaMgZn$} \cite{app}. A transition is observed from a
metallic regime (for a cubic approximant 1/1 ($a=14.2 \AA$)
$\rho_{4K}=58\mu\Omega cm$ and $\delta\rho(T)>0$) {\it to a ``quasiperiodic
regime"} for the approximant 2/1 (with parameter {\small $a=23\AA$}) with a
resistivity close to that of the icosehedral phase $\rho_{4K}=120\mu\Omega
cm$ and $\delta\rho(T)<0$).  This  suggests that a minimal size of the unit
cell is necessary to observe characteristic effects of quasiperiodic order.

\vspace{15pt}

\hspace{\parindent}The $AlPdRe$ alloy is even more resistive with a
behavior as a function of temperature different from the other
quasicrystals \cite{AvignonP}. Typically at 4K, the resistivity of the
alloy {\small $\rho(i-Al_{62,5}Pd_{22}Mn_{7,5})\sim 10.000\ \mu\Omega cm$}
whereas  {\small $\rho(i-Al_{70,5}Pd_{21}Re_{8,5})\sim 1. 500.000\
\mu\Omega cm$} which is of the same order of that of doped semiconductors
(note that Mn an Re belong to the same column of the Mendeelevtable and Mn
is in the 3d row whereas Re is in the 5d row).

\vspace{15pt}

\hspace{\parindent}So, the two icosahedral phases, defined by the same
long-range order, with close atomic composition, have nevertheless a
conductivity variation of several order of magnitude. The densities of
states at the Fermi level, for alloys like {\small $AlCuFe, AlPdMn,...$}
are  $\sim1/3$ of that of free electrons, whereas for {\small $AlPdRe$} it
is $\sim 1/10$ of that of free electrons. Consequently, the reduction of
conductivity between $i-AlPdMn$ and $i-AlPdRe$ is not only due to a
reduction of density of states, but depends also strongly on localization
issues. In figure {figRe1}, we present the resistivity variation of
$AlPdRe$ as a function of temperature (from \cite{gigith,Gignoux}).

between 

\vspace{15pt}

\hspace{\parindent}The resistivity at 4 Kelvin of the $AlPdRe$ alloy is
similar to that of resistive materials like doped semiconductors
\cite{Poon2}. So, it is legitimate to compare their respective behaviors as
a function of temperature in order to characterize the possible underlying
conduction mechanisms implied in $AlPdRe$ at low temperature. The
temperature dependence of {\small $\sigma(T)=1/\rho(T)$}  follows neither
an exponential law of  {\small $\exp(-E/k_{B}T)$} type
which is characteristic of thermally activated processes, nor
a {\small $\exp(-AT^{-1/4})$} law characteristic of variable range hopping
mechanisms at low temperature \cite{Mott}. On the contrary, the
conductivity of $AlPdRe$ follows on a large range of temperature $4-800K$,
a power law {\small $\sigma(T)\sim T^{\beta}$} with $1<\beta<1.5$
\cite{CB_RQ9} which remains unexplained.

\vspace{10pt}

\hspace{\parindent}As a comparison, we show the conductivity  of the
{\small $Al_{2}Ru$} alloy whose behavior is typical of thermally activated
process of semiconductors (represented in fig. {figPoon} with $\bullet$).
The behavior of $\sigma(T)$ is described by $exp(-\Delta/k_{B}T)$ (with
$\Delta=0.17 eV$ is the gap width)\cite{AvignonP}. The $i-AlPdRe$ alloy is
shown to follow $\sigma(T)\sim T^{\alpha}$.

\vspace{10pt}

\hspace{\parindent}The figure {figMIT} shows the correlation between
$\sigma_{4K}$ and $\sigma_{300K}$, which is an element to identify a
metal-insulator transition \cite{Berger2,Honda}. The position of  {\small
$AlPdRe$} on that plot is coherent with the special properties of this
alloy at low temperature.

\vspace{20pt}

\subsection{Quantum interferences in quasicrystals}

\vspace{20pt}

\hspace{\parindent}At low temperature, quantum interference effects (QIE)
have been clearly identified \cite{May1I,Take,Poon1},  for the class of
quasicrystals located on the metallic side, and over a large range of
temperature $[0.3K-100K]$  and magnetic field $[0-20]$ Tesla. These
phenomena of localization were expected in disordered systems of smaller
resistivity. However, the dependence $\delta\sigma(T)$ at low temperature
and $\delta\sigma(H)$ in magnetic field were analyzed convincingly by means
of the theories of quantum interference effects (weak localization,
electron-electron interaction..), even for the approximant cubic phase with
small parameter $a=12.33\AA$ de AlCuFeSi \cite{CB_RQ9}.

\vspace{10pt}

\hspace{\parindent}Concerning the $i-AlPdRe$ phase, the description of
$\sigma(T,H)$  by means of QIE is however not possible
\cite{CB_RQ9,gigith}. For these alloys, that seem to be on the insulating
side of a metal-insulator transition, it would be in fact surprising to
find for example weak localization effects.

\vspace{10pt}

\subsection{Conclusion on experimental results}

\vspace{20pt}

\hspace{\parindent}To conclude this experimental part, let us summarize the
main points concerning electronic transport properties. Many quasicrystals
and their approximant phases are typically composed by $60\%$ to $70\%$ of
aluminum which is a very good metal, but they appear to be very bad
conductors. In addition, they follow an ``Inverse Mathiessen rule" as a
function of temperature. We present in the figure {figFCP}, the resistivity
as a function of temperature for typical metallic alloys and quasicrystals
(from \cite{Berger3}).

\vspace{15pt}

\noindent
These systems are close to a metal-insulator transition. On the metallic
side, experiments reveal the importance of quantum interference effects and
suggest that the local order on $15-30\AA$ \cite{May2I} determine the
observed properties. On the insulating side, the power law $\sigma(T)\sim
T^{\alpha}$ for the conductivity of $AlPdRe$ on a wide range of temperature
is still unexplained.

\vspace{10pt}

\hspace{\parindent}To conclude, let us mention that other properties like
diamagnetism and small number of carriers (deduced from Hall effect) are
consistent with the tendency to localization observed in these systems.
Also, optical conductivity measurements show that there is no Drude peak
for icosahedral quasicrystals, which is again in contradiction with usual
characteristic of metals \cite{BergerAus}.

\vspace{20pt}

\section{Theory of electronic structure of quasicrystals}

\vspace{20pt}

\subsection{Localization and quasiperiodic order}

\vspace{10pt}

\subsubsection{Critical states and scaling of bands}

\vspace{20pt}

\hspace{\parindent}In order to study the spectral properties of independent
electrons in a quasiperiodic potential, one often uses a tight-binding
model where the vertices
 {\small ${\cal V}_{\sl T}$} of a given tiling are chosen as atomic sites
$\mid n\>$. The hamiltonian then writes :

\vspace{5pt}

$${\cal H}=\sum_{<n,m>\in {\cal V_{T}}}\mid n\>\ t_{nm} \ \langle m\mid + \
\sum_{n} \varepsilon_{n} |n\>\<n|$$

\vspace{5pt}

\noindent
where quasiperiodicity is introduced either geometrically (respective
positions of atomic sites), or through a modulation of energy sites or
hopping term on a periodic lattice. One of the main results concerning
electronic localization in Fibonacci chains is the power law behavior of
the envelope of the wave function {\small $|\psi_{N}|\sim  N^{-\alpha}$},
first studied by M. Kohmoto and B. Sutherland, and Ostlund et al.
\cite{Kho1I}, and refered to as {\it critical states}. Such a peculiar
localization is shown in figure {fig3} where an off-diagonal model is used
($\varepsilon_{n}=0 \ \forall n$) with {\small $t_{B}/t_{A}=\gamma$} the
measure of quasiperiodicity, and {\small $\alpha={\ln \gamma} / {\ln
{\tau}^{3}}$} exactly for $E=0$ \cite{Kho2f,Kho3}. For  2D-Penrose
lattices, the exponents are given typically by {\small $3/8<\alpha<5/8$},
depending on the eigenstates and the physical parameters \cite{SteinI}.

\vspace{10pt}

\hspace{\parindent}The properties of the eigenstate $\psi_{n}$ have to be
investigated through multifractal analysis \cite{Hent}. This eigenstate is
associated to a so-called 6-cycle, in relation to the transfer matrix
properties. Indeed, for this particular energy, starting from the first 6
elementary matrix ${\cal P}_{n=1,6}$ (with $n=F_{n}$ a Fibonacci number),
the algebraic properties of Fibonacci numbers allow to define a
renormalization scheme given by ${\cal P}_{n+6}={\cal P}_{n}$ which
constitute the 6-cycle. If $\psi_{0}$ and $\psi_{1}$  are taken as initial
conditions, it can be shown that the fractal structure of $\psi_{n}$ shown
in figure {fig3}
is determined by \cite{Kho3,HKubo3I} :

\vspace{5pt}

\begin{eqnarray}
\psi_{n}&=&{\gamma}^{2p}\psi_{0}\ \ \ \ \ \ \hbox{when} \ \ \
n=F_{3}+F_{6}+F_{9}+\ldots+F_{3\times (2p-1)}+F_{3\times 2p}\nonumber \\
\psi_{n}&=&-{\gamma}^{2p+1}\psi_{1}
 \ \ \hbox{when}\ \   n=1+F_{2}+F_{6}+\ldots+ F_{3\times 2p}+F_{3\times
(2p+1)} \nonumber
\end{eqnarray}

\vspace{5pt}

\noindent
with p a integer.

\vspace{10pt}

\hspace{\parindent}Concerning the spectral properties of quasiperiodic
hamiltonians in 1 dimension, J. Bellissard et al. have proven rigorously
the {\it gap labelling} theorem \cite{Bell2I,Bell4I}, which gives the value
of integrated density of states IDoS withineach gap of the electronic
structure. For a Fibonacci chain, the heights of the plateaux of IDoS are
given by {\small $j\tau/(\tau+1)$} mod1 where $j$ is an integer (see figure
figIDOS). C. Sire \cite{SirM1} has determined the width and the position of
gaps of periodic approximants of the Fibonacci chain in a perturbative
regime.

\vspace{10pt}

\hspace{\parindent}In higher dimension $>1$, several numerical studies on
discrete models (see for example studies on octogonal tiling \cite{Sir3I}
or Penrose tiling \cite{Fuji4I}) have complemented the description of
critical states. It is important to note that localization aspects can be
studied either directly in a pure quasiperiodic system, which becomes
rapidly difficult in higher dimensions, or by investigating the properties
of big approximants and focusing on the renormalization of properties from
one approximant to another.

\vspace{10pt}

\hspace{\parindent}In figure {figECP3} (from \cite{Fujipc}), the
localization of eigenstates is represented within a localized basis. The
projection of the given state in the unit cell of a Penrose approximant
allows to see three hierarchies of pentagona shift by an angle of $\pi/5$.
These states can be investigated through multifractal analysis \cite{Hent}
which lies in the evaluation of the moments of the probability
distribution, associated to spectral measure. If $\{|n\>\}$ is an
orthonormal basis, one can write :

\vspace{5pt}

$$\mu_{q}(\varepsilon_{k})=\frac{\sum_{n=1}^{N}|\<n|\Psi_{k}\>|^{2q}}{\strut
(\sum_{n=1}^{N}|\<n|\Psi_{k}\>|^{2})^{q}}\sim N^{-(q-1)D_{q}}$$

\vspace{5pt}

\noindent
where the exponents $D_{q}$ are known as the multifractal dimensions of the
spectrum.

\vspace{10pt}

\hspace{\parindent}Qualitatively, a critical state can be described as
follows : Suppose that a given state $\psi_{L}$ is mainly localized in a
region of characteristic length $L$. Then the Conway Theorem implies that a
similar region must exists at a distance $\leq 2L$. If $L$ is sufficiently
large, then both regions will be good candidate for a tunneling effect from
let's say the envelope of a given state is given by $\psi_{L}$ to
$\psi_{2L}=z\psi_{L}$, where a damping factor $z$ associated to the
probability amplitude of the event is introduced. Then the case $z=0$
corresponds to strictly localized states (Anderson localization), whereas
$|z|=1$ is the signature of extended states. For intermediate localization
cases, like those related to fractal eigenstates, one writes :

\vspace{5pt}

$$\psi_{L}\sim L^{ -\ln |z|/\ln 2} \sim  L^{-\alpha}$$

\vspace{5pt}

\noindent
in a perfect quasicrystal $|z|\neq 1$  and will generally
depend on the parameters of the considered model \cite{SirdosI}.

\vspace{10pt}

\hspace{\parindent}The nature of critical states can be also related
to the scaling properties of electronic dispersion relation of
approximants. Indeed, if we recall the argument of C. Sire
\cite{SirdosI}, one can focus on the scaling properties of bandwidth
of a series of approximants of a quasicrystal. If we considere an
initial cube of length $L$ in D dimensions with no restrictive
condition on the atomic order (periodic, disordered, quasiperiodic..),
then the spectrum  of the infinite periodic system of unit cell $L$
with $L^{D}$ atoms will be composed by $L^{D}$ bands. The typical
bandwidth is then related to the overlap between two
 states $\phi_{1}$ and $\phi_{2}$ localized in adjacent blocks
 of length $L$, and can be qualitatively linked with
 {\small $\Delta \varepsilon\sim |\<\phi_{1}|{\cal H}|\phi_{2}\>|$}.

\vspace{10pt}

\hspace{\parindent}For Bloch states with modulus {\small $|\phi(x)|\sim
1/{L}^{D/2}$} and with $t$ an average hopping amplitude from one site to
another, then {\small $\Delta \varepsilon\sim t/L$} and the mean velocity
writes $v\sim \Delta \varepsilon/L^{-1}\sim t$ which is, as expected,
independent of the arbitrary length $L$ chosen. The same argument for a
disordered cube involves the  localization length $\xi$ and the bandwidth
is then $\Delta \varepsilon\sim t L^{D-1}\exp(-L/\xi)$, which correspond
when $L\to\infty$  to a purely
discrete spectrum. For a system that will be dominated by algebraic
localization, i.e  $|\phi(x)|\sim1/L^{\alpha(x)}$ one finds a scaling
behavior of bandwidths  \cite{SirdosI} :

\vspace{5pt}

$$\Delta \varepsilon\sim \frac{t}{L^{\beta}}$$

\vspace{5pt}

\noindent
defined by an exponent $\beta>1$ related to the distribution of $\alpha$.
The mean group velocity as a function of the size $L$ is then

\vspace{5pt}

$$ v \sim \frac{\Delta \varepsilon}{L^{-1}} \sim \frac{t}{L^{\beta-1}}$$

\vspace{5pt}

\noindent
which goes to zero when $L\to\infty$. The $\beta$ exponents have been
studied for 1D and 2D systems \cite{Fuji4I} and confirm this general
argument.

\vspace{20pt}

\subsubsection{Quantum diffusion in perfect quasiperiodic structures}

\vspace{20pt}

\hspace{\parindent}Since the band velocity depends on the length scale, one
expects non ballistic propagation in a perfect quasiperiodic structure.
The dynamic of an initially localized wave packet (such that {\small
$\psi_{n}(t=0)=\delta_{n,n_{0}}$}) is governed  by the time dependent
Schr\"{o}dinger equation :

\vspace{5pt}

$$i\hbar \frac{d\psi_{n}}{dt}
=-t(\psi_{n+1}-\psi_{n-1})+\varepsilon_{n}\psi_{n}$$

\vspace{5pt}

\noindent
H. Hiramoto and S. Abe \cite{Abe} have studied, in the diagonal model
($t_{n}=t\ \forall \ n$), the diffusion properties of such wave packets in
Fibonacci chains
and have shown that the mean square displacement was defined by an abnormal
regime of propagation :

\vspace{5pt}

$$\sqrt{\langle (\Delta x)^{2} \rangle}=\sqrt{\sum_{n}(n-n_{0})^{2}\mid
\psi_{n}(t)\mid^{2} }\sim t^{\gamma}$$

\vspace{10pt}

\noindent
where the $\gamma$ coefficients vary between the typical values of
localized states ($\gamma=0$ with {\small $\sqrt{\< {(\Delta x)}^{2} \>}$}
which remains finite when {\small $t\rightarrow\infty$}), and ballistic
regimes  $\gamma=1$ for extended states. This abnormal regime of
propagation of wave packets is specific to the correlations of a
quasiperiodic potential.

\vspace{10pt}

\hspace{\parindent}The quantum diffusion properties are characterized
by the whole distribution of moments associated to the spread of a
given wave packet {\small $\<x^{q}\>(t)={t}^{q\sigma_{q}}$}. Recently,
F. Pi\'echon, using a perturbative renormalization group treatment
\cite{Piech1} has investigated analytically the diffusion exponent of
the wave packets in Fibonacci chains. In particular, he has shown that
there exists a fraction of sites  from which a quantum state will
spread with exponents  $\sigma_{q}=\ln \tilde{z} \ln{\tau}^{-3}$ whereas
the complementary remaining sites are associated with  $\sigma_{q}=\ln
z  \ln {\tau}^{-2}$ (where $\tilde{z}, z$ are the perturbative parameters
related to the quasiperiodic potential). The main result of his study
is the fact that diffusion exponents of wave packets are directly
related to exponents of the spectral measure. We also note that for
systems in higher dimension, G. Jona-Lasinio et al. \cite{Lasinio}
have rigorously shown that a propagation of wave packets in
hierarchical potential,
 through tunneling over arbitrary large scales is related to anomalous
diffusion.

\vspace{20pt}

\subsubsection{Landauer Resistance in quasiperiodic systems}

\vspace{20pt}

\hspace{\parindent}By means of the formalism of transfer matrix, M. Kohmoto
et al. \cite{Kho2A} have conjectured an algebraic increase of Landauer
resistance. They have shown analytically, for energies corresponding to
Q-cycles, that {\small $\rho_{N}\leq \ \rho_{0}\  N^{\alpha}$}. The authors
extend heuristically their results to chaotic but bounded orbits, and
rigorous mathematical approaches have demonstrated \cite{Ray} that for all
$E$ belonging to the spectrum $\sigma({\cal H})$ (singular continuous) but
constituting a zero measure ensemble, the exacts solutions of the
Schr\"{o}dinger equation {\small ${\cal H}\psi=E\psi$}, as well as the
Landauer resistance were bounded by polynomials.

\vspace{20pt}

\hspace{\parindent}The works of Tsunutsegu have generalized these power law
dependences of landauer resistance for two dimensional Penrose lattices
\cite{Tsu}. The Landauer conductance follows a general power law  {\small
$g(L)\sim {L}^{\alpha}$} whose typical exponents are given in figure
{Tsufig}. The system studied consists of a finite part of a Penrose lattice
which is connected to conducting leads of finite width M.

\vspace{20pt}

\subsection{Electronic structure of quasicrystalline materials}

\vspace{30pt}

\hspace{\parindent}To investigate electronic properties of more realistic
materials, one can use ab-initio methods and study the general
characteristics of electronic structure of approximant phases of
quasicrystals. From ab-initio calculations, the main features can be
summarized by the existence of a deep pseudo-gap at the Fermi level
\cite{Fuji1I}, and by an overall structure composed of a high concentration
of bands with small dispersion (see figure {figSTB2}) (from ref.
\cite{Fuji2I,guyth}).

\vspace{10pt}

\hspace{\parindent}The presence of a deep pseudo-gap at $E_{F}$ is familiar
to the alloys of Hume-Rothery type. This point has been experimentally
confirmed by X-ray emission or absorption spectroscopy for icosahedral
phases \cite{Esther}. The specific heat measures give also a weak TDoS at
the Fermi level, of order $\sim 1/3$  of free electrons for icosahedral
phases of $AlCuFe$, $AlLiCu$ \cite{Esther} (see fig. {figPseudga}), and of
$\sim 1/10$ for $i-AlCuRu$ \cite{D2} and $i-AlPdRe $ \cite{Poon2}. The
pseudo-gap is also observed experimentally for approximant phases
\cite{D4}.

\vspace{10pt}

\subsubsection{Hume-Rothery mechanism and pseudo-gap at the Fermi level}

\vspace{10pt}

\hspace{\parindent}Let us recall briefly the physical mechanism at the
origin of pseudo-gap in metallic alloys. If we consider nearly free
electrons, that is free electrons diffused by a weak pseudopotential $V$,
the hamiltonian ${\cal H}$  of the system reads : {\small ${\cal H} = {\cal
H}_{0} + V$} with {\small ${\cal H}_{0}=\hbar^{2}k^{2}/2m$} the kinetic
energy of free electrons. Within the local approximation of
pseudopotentials, $V$ is given by :

\vspace{5pt}

$$V({\bf r})=\sum_{\bf K}V_{\bf K}\exp(i{\bf K.r})$$

\vspace{5pt}

\noindent
where ${\bf K}$ is a vector of reciprocal space. The
Fourier coefficient
  $V_{\bf K}$ will couple the plane waves  $|{\bf k}\>$ and $|{\bf
k-K}\>$. The kinetic energies of plane waves $|{\bf k}\>$ and  $|{\bf
k-K}\>$ are close as soon as ${\bf k}$ is close to the Bragg
plan associated to  ${\bf K}$.
The mixing between $|{\bf k}\>$ and  $|{\bf k-K}\>$
 is strong near this Bragg plane. Furthermore, the stronger
 $|V_{\bf K}|$, the stronger is this mixing.

\vspace{5pt}

\hspace{\parindent}Let us now consider the effect of one Bragg plane,
associated to  ${\bf K}$, on a plane wave $|{\bf
k}\>$. As soon as
   ${\bf k}$ is sufficiently close to the Bragg plane, we neglect the
contribution of the other Bragg planes by considering only  $V_{\bf
K}$ which couple states  $|{\bf k}\>$ and
 $|{\bf k-K}\>$ (two bands model), then the hamiltonian matrix is :

\vspace{5pt}

\begin{center}
\begin{eqnarray}
\left(
\bea{cc}
\frac{\hbar^{2}}{2m} {\bf k}^{2}&  V^{*}_{\bf K}\\
V_{\bf K} & \frac{\hbar^{2}}{2m}({\bf k-K})^{2} \\
\ea
\right)
\end{eqnarray}
\end{center}

\vspace{5pt}

\noindent
The dispersion relations of such a system are represented schematically in
figure {figEK}. The effect leads to the formation of bonding and
anti-bonding states within the free electron band, and this leads to the
apparition of two Van-hove singularities in the density of states which for
increasing energies form a peak followed by a pseudo-gap. In the weak
potential limit, it can be shown that the minimum of the pseudo-gap lies at
energy {\small $E_{0}({\bf K}/2)+|V_{\bf K}|$}. The point is that the
structure is stabilized when the Fermi level lies in the pseudo-gap
(Hume-Rothery rule \cite{HR}).

\vspace{10pt}

\hspace{\parindent}Soon after their discovery, quasicrystals and their
approximant phases have been rapidly classified as Hume-Rothery alloys
\cite{Friedel}. In particular,
J. Friedel and F. D\'noyer \cite{Fried3} have shown that the
icosahedral phase of AlLiCu was stabilized by a strong interaction
between a Fermi surface of electrons and a ``pseudo Brillouin zone"
(\cite{Jones}). The latter was constructed from the Bragg planes
 corresponding to the brightest peaks of the diffraction pattern.
In figure figZ, we present two examples of pseudo-Brillouin-zone of
the icosahedral.
 The  stabilization of  i-AlCuFe involves the pseudo-zone  A,
whereas for i-AlLiCu, the pseudo-zone B is involved.

\vspace{20pt}

\subsubsection{Spiky structure of the density of states and localization}

\vspace{20pt}

\hspace{\parindent}The spiky structure of the TDoS observed in figure
figSTB2 corresponds
 to a high and inhomogeneous concentration of bands with small
dispersion, whose first consequence as far as transport is concerned
is a small velocity at Fermi energy {\small $v(E_{F})=(\partial E
/\partial k)|_{E=E_{F}}$}. It is interesting to note that even for
simple models, where quasiperiodic order is introduced, for example
through modulation of energy sites, the TDoS (see
figPdg) also reveal a spiky structure (from \cite{SirdosI}).

\vspace{10pt}

\hspace{\parindent}Recently, T. Fujiwara et al. \cite{Fuji5I} have
investigated the electronic properties of localization for 3D structural
models of decagonal approximant phases {\small $Al_{66}Cu_{30}Co_{14}$}.
Developing the eigenstates within a basis of {\bf spd} orbitals , {\small
$\chi_{{\bf R},l}({\bf r})=\<{\bf r}|{\bf R},l=0,1,2\>$}, {\small
$\Psi_{k}({\bf r})=\sum_{ {\bf R},l=0,1,2} \alpha_{{\bf R},l} \chi_{{\bf
R},l}({\bf r})$ }, they have shown that the inverse participation ratio was
described by a power law as a function of the approximant size. For a given
Fermi level, they have evaluated :

\vspace{10pt}

$$P_{\Psi}=1/\mu_{2}(\varepsilon(k))=\bigl(\sum_{{\bf R},l} \mid
\alpha_{{\bf R},l} {\mid}^{2}\ \bigr)^{2}/ \sum_{{\bf R},l} \mid
\alpha_{{\bf R},l}{\mid}^{4}$$

\vspace{10pt}

\noindent
and shown that {\small $P_{\<\Psi\>}\sim N^{\nu}$}, for an average over
states with energy close to $E_{F}$, and as a function of the number $N$ of
atoms per unit cell. This quantity, which gives the number of sites which
contribute significantly to the state $|\Psi\>$ for a given energy (see
\cite{Fuji5I}),
is hence defined by a scaling behavior of a state whose distribution of
weight in a large unit cell is inhomogeneous (not uniformly delocalized).
This power law on a realistic model of approximant suggests that
localization properties of approximants are similar to those of
quasicrystals.

\vspace{10pt}

\hspace{\parindent}These aspects of band structure allow to evaluate the
behavior of Boltzmann conductivity in the relaxation time approximation
RTA. The L.M.T.O. results obtained by T. Fujiwara et al. \cite{Fuji1I},
have confirmed that conductivity is very weak, and more concretely they get
{\small $\sigma_{DC}\sim 10-150 (\Omega cm)^{-1}$} for the {\small
$\alpha-$AlMn(Si)} phase  (cubic lattice of Mackay icosahedra), which are
anomalously low for metallic alloys. Their adjustable parameter is the
scattering time $\tau$ for which they take standard values {\small $ \sim
10^{-14,-15} s$}. In addition, their study of the decagonal phase {\small
$d-AlCuCo$} agrees with {\it the observed experimental anisotropy
\cite{anisoI} between quasiperiodic and periodic directions}({\small
$\sigma_{DC}\sim 3000 (\Omega cm)^{-1}$})for the quasiperiodic one, whereas
for {\it periodic} ({\small $\sigma_{DC}\sim 15000 (\Omega cm)^{-1}$})
\cite{Fuji2I}. Another analytical approach of a Bloch-Boltzmann type by
S.E. Burkov et al. \cite{BurkovI} also confirms the general tendency of
weak conductivity in quasicrystals.

\vspace{10pt}

\hspace{\parindent}Recently, we have proposed a model to analyze the role
of atomic clusters (see section 2.3) in electronic structure of
quasicrystalline materials \cite{guyrq9,guyprl}. The model is a cluster of
transition metal atoms in a metallic matrix. We analyzed the modifications
of density of states due to multiple scattering effects.The essential
result is that the spiky peaks of the TDoS observed from ab-initio
calculations, and that seem to have been observed experimentally
\cite{Davydov}, could be the signature of states preferentially localized
around structural cluster. We propose a generalization of the concept of
virtual bound state to {\it cluster virtual bound states}.

\vspace{10pt}

\hspace{\parindent}We show in the figures {figag1}, different cluster
virtual bound states according to the geometry of the structural entity
(icosahedra, dodecahedra...) \cite{guyrq9}. We note that, according to our
results, the existence of spiky structure is favoured for icosahedral
cluster
for which lifetime of an electron will be longuest. Also, figure {figag2}
shows how a cluster of clusters induce again more spiky structure in the
density of states at the scale of a more complex cluster.

\vspace{30pt}

\section{Theoretical approaches to electronic transport in quasicrystals}

\vspace{20pt}

\hspace{\parindent}A semi-classical Bloch-Boltzmann description of
transport in quasicrystals seems unsufficient to take into account  most of
the aspects due to the special algebraic localization of these materials.
Some specific transport mechanisms have been proposed to explain results
like the inverse Mathiessen rule, or the temperature and defects influence
on the conductivity, that need to go beyond a Bloch-Boltzmann analysis.

\vspace{10pt}

\hspace{\parindent}In particular, the possibility of two different
unconventional transport mechanisms specific of these materials has been
proposed \cite{May1I,Fuji2I,EsMay}. Transport could be dominated, for short
relaxation times $\tau$ (defined by disorder) by hopping between ``critical
localized states", whereas for long time $\tau$ the regime could be
dominated by non-ballistic propagation of wave packets between two
scattering events. We present briefly these two mechanisms in the
following.

\vspace{10pt}

\subsection{Non-ballistic propagation between two collisions events}

\vspace{20pt}

\hspace{\parindent}A possible interpretation of conductivity in
quasicrystals \cite{May1I,EsMay}, takes as a starting point the numerical
results on anomalous diffusion of wave packets in perfect quasiperiodic
structures \cite{Sir4I} (see section 4.1.1). Indeed for sufficiently long
collision time $\tau$ (due to disorder), the propagation of wave packet
follows typically a law such that $L(t)=At^{\beta}$ with $\beta < 1$ which
defines a sub-ballistic regime with  $A$, a constant which does not depend
explicitely on time. It is then possible to estimate the conductivity by
means of Einstein formula {\small $\sigma={e}^{2}N(E_{F})D(\tau)$}.  The
diffusivity is given as a function of  $\tau$ by {\small
$D(\tau)=L^{2}(\tau)/3\tau$} and {\small $N(E_{F})$} is the density of
states at Fermi energy. The explicit dependence of conductivity with $\tau$
thus reads :

\vspace{5pt}

$$\sigma(\tau)\sim A {\tau }^{2\beta-1}$$

\vspace{10pt}

\hspace{\parindent}In consequence, provided that $\beta\leq 0.5$,
conductivity will decrease with $\tau$, in other words with the diminution
of defects. In the limit where $\beta\sim 0$ we will get $\sigma \sim
1/\tau$, which will be in agreement with the inverse Mathiessen rule that
characterize quasicrystals \cite{Klein2} (i.e a general form of the
conductivity  {\small $\sigma (T)=\sigma_{4K}+\delta\sigma (T)$}). Such
values of exponents have been obtained in some regimes on octagonal tilings
\cite{Sir4I}, but they correspond to strong quasiperiodic potential.

\vspace{10pt}

\hspace{\parindent}A similar argument is due to C. Sire \cite{Sir4I}, which
states that for a given mean free path $l_{pm}=L$, the qualitative behavior
of a quasicrystal and an approximant of unit cell $L$ will be the same. So,
as for an approximant, we can write the mean velocity {\small
$v_{F}=v_{F}(L)\sim {L}^{-\alpha}$} with {\small $L=v_{F}\tau $}, the
diffusivity then reads :

\vspace{5pt}

$$D(L)\sim {v}_{F}^{2}(L)\tau\sim {L}^{1-\alpha} \ \ \hbox{which is
equivalent to the first argument}$$

\vspace{10pt}

\hspace{\parindent}J. Bellissard and H. Schulz-Baldes \cite{Bell3I} have
shown recently, by a mathematical study of Kubo-Greenwood formula in a
model of quasicrystals, that for long relaxation time,
$\tau\longrightarrow\infty $, the transport law gives {\small $\sigma \sim
{\tau}^{2\beta-1}$} where the $\beta$ exponent is linked to anomalous
quantum diffusion regime.

\vspace{20pt}

\subsection{Interband transition mechanisms and hopping transport}

\vspace{30pt}

\hspace{\parindent}If one considers periodic hamiltonians whose eigenstates
are inhomogeneously distributed in a large unit cell, an interband
transition mechanism  could allow a new kind of propagation modes
\cite{May1I}. Indeed, the eigenstates of the perfect structure must have
the generic Bloch form {\small $\Psi_{nk}(r)=u_{nk}(r){e}^{ikr}$} with $n$
a band index, $k$ a wave vector and $u_{nk}$ a periodic function defined in
the unit cell (we suppose that the coefficients {\small $|u_{nk}({\bf
r})|$} have an unequal distribution of weight through the large unit cell,
as we schematize in fig. {figEC1}. Then collision events can induce
transitions from $\Psi_{nk}(r)\longrightarrow \Psi_{n'q}(r)$, which will
correspond to a charge displacement in real space. This phenomenon is not
taken into account in a Bloch-Boltzmann description.

\vspace{10pt}

\hspace{\parindent}The comparison between the contributions of this hopping
mechanism and of the classical metallic conductivity (i.e. propagation of
charge between two scattering events) could be made as follows. Suppose
that
$\Delta\varepsilon$ is a typical bandwidth at a given scale $L$ and $\tau$
a finite lifetime induced by disorder. Then, within this framework of
hopping mechanism, the dominant modes of transport will be given by the
comparison between the hopping frequency  $\Delta\varepsilon/\hbar$ due to
the quasiperiodic potential and the frequency $1/\tau$ induced by disorder.
Consequently, the latter mechanism will dominate provided that
$1/\tau>\Delta\varepsilon/\hbar$. This could be realised even for small
disorder in quasicrystalline materials, given that band dispersion is very
weak.

\vspace{10pt}

\hspace{\parindent}In this regime of interband transition, the hopping
length will not depend anymore on collision time so that diffusivity will
write $D=L^{2}(\tau)/3\tau \sim 1/\tau$. This explains qualitatively the
inverse Mathiessen rule, as well as the  variation of conductivity with
improvement of structural order $\sigma\sim 1/\tau$. This interband
mechanism has been also proposed by T. Fujiwara \cite{Fuji2I} as a natural
consequence of band structure and from numerical results on 2D Penrose
lattices \cite{Fuji3I,Fuji2I} where random phasons defects are introduced
in the lattices. We refer also to our studies of Landauer resistance of
quasiperiodic chains \cite{Ro3} and on quantum networks \cite{Ro4}, for
some exact results on the role of phasons.  C. Janot \cite{Janot2} has also
proposed a multiscale hopping mechanism in relation to hierarchical aspects
of structural models.

\vspace{20pt}

\subsection{Quantum interference and correlated disorder}

\vspace{30pt}

\hspace{\parindent}A first argument to justify the observation of quantum
interference effects (weak localization, electron-electron interaction..)
in quasicrystals is to assume that the general scaling theory of
localization is applicable as soon as electrons are in a diffusive regime.
The phase coherence is preserved only up to an inelastic length imposed by
finite temperature or magnetic field.

\vspace{10pt}

\hspace{\parindent}From several experiments \cite{May2I}, an estimation of
the elastic mean free path at zero temperature $l_{pm}\sim 20-30\AA$ for
$AlCuFe, AlPdMn..$ alloys has been proposed. Then the minimal Mott
conductivity is  $\sigma_{min}\sim 200\times (3/l_{pm})\sim 20-30 (\Omega
cm)^{-1}$, with a ratio of $\sigma_{exp}/\sigma_{min}\sim
(100-150)/(20-30)\gg 1$. This suggests that quantum interference do not
dominate the conduction regime for these systems. But, if we now apply the
same procedure for $AlPdRe$, assuming the same mean free path, $l_{pm}\sim
20-30\AA$, we get  {\small $\sigma_{exp}/\sigma_{min}\sim 1/(20-30)\ll 1$}.
In conclusion the regime is now dominated by quantum interference at the
origin of a metal-insulator transition. It could be that the quasiperiodic
potential is stronger in $AlPdRe$, leading to a smaller conductance
$g<g_{c}$ for a cube with size equal to the mean free path ($g_{c}$ is the
universal critical conductance).

\vspace{10pt}

\hspace{\parindent}There is another way to conceive localization in
quasicrystals, taking as a starting point the notion of correlated
disorder. Indeed, one can expect that disorder in a quasicrystal preserves
strong correlations between diffusive centers whereas classical theories of
localization assume a completely random distribution of scatterers.

\vspace{10pt}

\hspace{\parindent}A first example in this context, is a calculation of
quantum corrections, on the metallic side of an Anderson metal-insulator
transition, for a disordered binary alloy with local and short range
interaction between impurities \cite{Beal}. M.T. B\'eal-Monod and G. Forgac
have evaluated the quantum corrections of the electronic conductivity,
introducing conditionnal probability in the spatial distribution of
diffusion centers, and have shown, for example, that a repulsive first
neigbhours interaction will decrease conductivity with regards to the
uncorrelated case.

\vspace{10pt}

\hspace{\parindent}In the context of mesoscopic physics \cite{ALW} and
phase coherent effects, M. B. Hastings et al. \cite{Hast}
 have analyzed the effects of real space symmetries of the diffusion
potential. They have shown explicitely how symmetry constraints imposed on
disorder, influence diagrammatic treatments.

\vspace{10pt}

\hspace{\parindent}One of us, has investigated the effect of some
particular phason defects in Fibonacci chains and quantum networks Landauer
resistance \cite{Ro3,Ro4}. These studies show how correlations of disorder
induced by this kind of defects can lead to specific behavior of
conductivity.

\vspace{30pt}

\subsection{Kubo-Greenwood conductivity in quasiperiodic systems with disorder}

\vspace{30pt}

\hspace{\parindent}From ab-initio calculations of band structure, it has
been shown that Bloch-Boltzmann conductivity in pure approximants was weak,
in good agreement with experiments. However, in order to study carefully
the transport phenomena in quasicrystals, we need to go beyond the
semi-classical approximations. In particular, the heuristic arguments given
above suggest that the Bloch-Boltzmann
picture of ballistic propagation between scattering events is not applicable.

\vspace{10pt}

\hspace{\parindent}Thus, a natural starting point for a study of transport
is given by the Kubo-Greenwood linear response theory of transport
coefficient \cite{Kubo,Greenwood}, which makes no assumptions on the nature
of states.  At zero temperature and frequency one gets for a  Fermi level
$E_{F}$ :

\vspace{10pt}

$$\sigma_{DC}(E_{F})=\frac{2\pi {e}^{2}\hbar}{\Omega}
\hbox{Tr}[\hat{V}_{x}\delta(E_{F}-{\cal H})\hat{V}_{x}\delta(E_{F}-{\cal H})]
$$

\vspace{20pt}

\hspace{\parindent}Recently, we have developed \cite{Ro0,Ro1} a numerical
method based on the theory of orthogonal polynomials to evaluate directly
this formula. We have investigated the relation between quantum diffusion
of wave packets and conductivity at Fermi energy for a 3D quasiperiodic
system. We used a diagonal tight-binding model where quasiperiodicity was
introduced through a modulation of site energies. As propagation in pure
quasiperiodic systems is not diffusive, we have also introduced a
disordered potential which established a diffusive regime at long time of
wave packets diffusion.

\vspace{20pt}

\hspace{\parindent}We clearly show that transport regime deviates from the
prediction of a Bloch-Boltzmann approach which is applicable to metals in
the limit of weak disorder. The Bloch-Boltzmann theory predicts that the
conductivity varies like $\sigma=\sigma_{0}(V_{0}/V_{dis.})^2$. Thus, when
the strength of the disorder potential is multiplied by $\sqrt{2}$, the
conductivity must be divided by two. It is clear from figures {figcond1}
and {figcond2} that this prediction is not verified. The conductivity is
less sensitive to disorder than what Bloch-Boltzmann theory predicts.
Another conclusion is also that for sufficiently strong quasiperiodic
potential, the conductivity is nearly insensitive to disorder in the
regions of pseudo-gaps. This result is quite interesting because the Fermi
level of quasicrystalline materials has been shown to be located within a
pseudo-gap region. It suggests that at these energies, the quasiperiodic
potential acts like a strongly disordered potential and thus the
conductivity is nearly independant to a small additional source of
scattering \cite{Ro1}.

\newpage

\section{Conclusion}

\vspace{20pt}

\hspace{\parindent}Experiments show that quasicrystals present remarkable
electronic transport properties. In particular, they approach a
metal-insulator transition when the quasiperiodic order is improved. Within
the one electron scheme, the different theoretical approaches suggest an
intermediate type of algebraic localization between purely extended states
and exponentially localized states. Several schemes have been proposed to
explain original transport properties, among which, non-ballistic quantum
diffusion for long relaxation times $\tau$ or hopping mechanisms induced by
interband transitions at shorter $\tau$. We note however that the role of
electron-electron interactions has not been discussed up to now although
they can play an important role close to a metal insulator transition. More
precisely, the quasiperiodic potential tends to localize electrons and
under these circumstances, the electron-electron interaction could thus
lead to a Coulomb-gap (or pseudogap) even without disorder. Electron
interactions could lead to changes in the band structure close to the Fermi
level that cannot be described by the usual treatment of correlations used
in ab-initio calculations (i.e the local density approximation (L.D.A.)).
These changes can be sensitive to defects, scattering by phonons or
variation of the Fermi Dirac distribution with temperature. This could
deeply affect the transport properties of quasicrystals \cite{ICQ6}.

\vspace{20pt}

\section{Acknowledgments}
We acknowledge many fruitful discussions with C. Sire, C. Berger, T.
Fujiwara, J. Bellissard, F. Pi\'echon and N. Ashcroft.

\newpage

\noindent
Figure Captions:

\vspace{20pt}

\noindent
Figure 1 : Construction of the Fibonacci chain and an approximant one
through  cut and project algorithm.

\vspace{20pt}

\noindent
Figure 2 :  Penrose tiling of a 2D space.

\vspace{20pt}

\noindent
Figure 3 :  Schematic diffraction pattern of a Penrose tiling.

\vspace{20pt}

\noindent
Figure 4 :  Successive atomic layers of a  pseudo-Mackay icosahedron (PMI).
1: centered cube , 2: icosahedron, 3: icosidodecahedron, 4: PMI. The model
of C. Janot   distinguishes two PMI according to their atomic composition.
A first family (PMI-A) is defined with 6 manganese atoms plus 6 paladium on
the icosahedron, with the other sites occupied by aluminum, while a second
family (PMI-T) has 20 paladium atoms among the 30 available sites on the
icosidodecahedron, with aluminum atoms for the other sites

\vspace{20pt}

\noindent
Figure 5 : Example of a plane section of the 3D structure, where the
circles correspond to the equatorial sections of a PMI. A ${\tau}^{3}$
inflation is necessary to find a PMI of PMI

\vspace{20pt}

\noindent
Figure 6 : Electronic conductivity for differents quasicrystalline phases
as a function of temperature and structural quality (courtesy of C.
Berger).

\vspace{20pt}

\noindent
Figure 7 : Comparison of the $\sigma(T)$  between icosahedral phases and an
approximant  alloy of unit cell parameter $a=12.33 \AA$

\vspace{20pt}

\noindent
Figure 8 : Strong variation of the resistivity of AlPdRe as a function of
temperature

\vspace{20pt}

\noindent
Figure 9: $\sigma(T)$ for $i-Al_{70}Pd_{20}Re_{10}$ and $Al_{2}Ru (\bullet)$

\vspace{20pt}

\noindent
Figure 10 : $\sigma_{4K}$ as a function of $\sigma_{300K}$ for different
icosahedral phases

\vspace{20pt}

\noindent
Figure 11: Magnetoconductivity $\Delta\sigma(H)/\sigma_{0}\
\hbox{for}\ i-AlCuFe$ up to 20 Tesla

\vspace{20pt}

\noindent
Figure 12: General behaviors of conductivity for quasicrystals compared to
metallic alloys (courtesy of C. Berger).

\vspace{20pt}

\noindent
Figure 13: Fractal property of eigenstates localization (through $\mid
\psi_{n}\mid$) in a pure Fibonacci chain

\vspace{20pt}

\noindent
Figure 14: Integrated density of states $IDoS(E)$ for the Fibonacci chain
which appears to be a increasing continuous function of energy, not
differentiable.

\vspace{20pt}

\noindent
Figure 15: Projection of an eigenstate in the unit cell of a Penrose
approximant (N=9349 sites). One can distinguish three hierarchical
pentagonal entities formed by inhomogeneous distribution of amplitudes. The
grey and black sites represent more than $90\%$ of the total weight of the
considered state.

\vspace{20pt}

\noindent
Figure 16: Conductance $g(L)$ for $E=-3.4$ for different widths M in the
semi-periodic direction

\vspace{20pt}

\noindent
Figure 17: Total densities of states (TDoS)  of alloys with close
composition : (A) approximant 1/1 $i-Al_{62.5}Cu_{25}Fe_{12.5}$ 128
atoms/unit cell, (B) non-approximant  $\omega-Al_{7}Cu_{2}Fe$, 40
atoms/unit cell.

\vspace{20pt}

\noindent
Figure 18: Dispersion curves $E(k)$ along the main directions of the
reciprocal space.

\vspace{20pt}

\noindent
Figure 19: Partial density of states measured by X-ray emission or
absorption spectroscopy (from ref. \protect\cite{Esther}). (a) pure Al, (b)
$\omega-Al_{7}Cu_{2}Fe$, (c) rhombohedral approximant
$Al_{62.5}Cu_{26.5}Fe_{11}$, (d) icosahedral phase
$Al_{62}Cu_{25.5}Fe_{12.5}$.

\vspace{20pt}

\noindent
Figure 20 : Effect of Bragg diffraction on electronic bands. (a): $V_{K}=0,
(b) V_{K}\neq 0$.

\vspace{20pt}

\noindent
Figure 21: Representation of a Pseudo-gap.

\newpage

\vspace{20pt}

\noindent
Figure 22: Examples of pseudo-Brillouin-zones of the icosahedral phase. A:
42 (30+12) facets (main pseudo-zone for $AlCuFe, ALPdMn$; B: 60 facets
(main pseudo-zone for $AlCuLi$). The arrows are issued from the peaks which
together with all equivalent peaks (by the icosahedral symmetry) define the
facets of the pseudo-zone.

\vspace{20pt}

\noindent
Figure 23: TDoS of two cubic approximants with L=13 et L=144 sites per unit
cell (courtesy of C. Sire).

\vspace{20pt}

\noindent
Figure 24: Cluster virtual bound states. (b): Tetrahedron, (c): Cube, (d):
Icosahedron, (e): Dodecahedron. (a): Friedel virtual bound state.

\vspace{20pt}

\noindent
Figure 25: Cluster virtual bound states on a icosahedron (b) and on a
icosahedron of icosahedron (c).

\vspace{20pt}

\noindent
Figure 26: Schematic illustration of a critical state.

\vspace{20pt}

\noindent
Figure 27: Schematic representation of an interband transition. $D_{g}$ is
the  displacement of the gravity center of the wave packet.

\vspace{20pt}

\noindent
Figure 28: Electronic conductivity as a function of Fermi energy for a
quasiperiodic potential defined on a cubic lattice with nearest-neighbour
hopping $t=1$. The potential is given by on-site energies
$\varepsilon_{ijk}=\varepsilon_{i}+\varepsilon_{j}+\varepsilon_{k}+V_{dis.}$
, with $\varepsilon_{i}=\pm V_{qp}=0.7$ according to a Fibonacci sequence,
and $\varepsilon_{dis.}$ is randomly distributed between
$[-V_{dis.}/2,V_{dis.}/2]$ $V_{dis.}=2: (a), V_{dis.}=2\protect\sqrt{2}:
(b)$). The insert shows the average over sites j of
$D_{j}(t)={(r(t)-r_{j})}^{2}/t$ for a wave packet initially localized in j.
This shows the onset of a diffusive regime for long time ($D_{j}(t)=cste$).

\vspace{20pt}

\noindent
Figure 29: Electronic conductivity as a function of Fermi energy for the
model defined on figure 29, but for a quasiperiodic potential,
$V_{qp}=0.9$, and $V_{dis.}=2: (a), V_{dis.}=2 \protect\sqrt{2}: (b)$.

\end{document}